\documentclass[%
 aip,
 sd,%
 amsmath,amssymb,
 reprint,%
]{revtex4-1}

\usepackage{graphicx}
\usepackage{dcolumn}
\usepackage{bm}
\usepackage{color}
\usepackage{threeparttable}
\usepackage{tabularx}
\usepackage{threeparttable}
\usepackage{lineno}
\usepackage{multirow}
\usepackage{amssymb}

\begin{document}

\preprint{AIP/123-QED}

\title{The Effect of Cesium Carbonate on PCBM Aggregation in Films}

\author{William R. Lindemann}
 \affiliation{Ames Laboratory, and the Department of Physics and Astronomy, Iowa State University, Ames Iowa 50011, USA}
\author{Wenjie Wang}
 \affiliation{Ames Laboratory, and the Department of Physics and Astronomy, Iowa State University, Ames Iowa 50011, USA}
\author{Fadzai Fungura}%
\affiliation{The Microelectronics Research Center,  and the Department of Electrical \& Computer Engineering, Iowa State University, Ames, Iowa 50011, USA
}%
\author{Joseph Shinar}%
 \affiliation{Ames Laboratory, and the Department of Physics and Astronomy, Iowa State University, Ames Iowa 50011, USA}
\author{Ruth Shinar}%
 \email{rshinar@iastate.edu}
\affiliation{The Microelectronics Research Center,  and the Department of Electrical \& Computer Engineering, Iowa State University, Ames, Iowa 50011, USA
}%
\author{David Vaknin}
 \email{vaknin@ameslab.gov}
 \affiliation{Ames Laboratory, and the Department of Physics and Astronomy, Iowa State University, Ames Iowa 50011, USA}

\date{\today}

\begin{abstract}
Surface-pressure versus molecular area isotherms, X-ray reflectivity and X-ray near-total reflection fluorescence were used to study the properties of 1-(3-methoxycarbonyl)propyl-1-phenyl[6,6]C$_{61}$ (PCBM)  that was pre-mixed with cesium carbonate and spread as a film at the air-water interface.  The pre-mixed PCBM with cesium carbonate demonstrated a strikingly strong effect on the organization of the film. Whereas films formed from pure PCBM solution were rough due to strong inter-molecular interactions, the films formed from the mixture were much smoother.  This indicates that the cesium carbonate moderates the inter-molecular interactions among PCBM molecules, hinting that the cesium diffusion observed in inverted organic photovoltaic structures and the likely ensuing ionic Cs-PCBM interaction decreases aggregation tendency of PCBM. This implies that the use of cesium salts affects the morphology of the organic layer and consequently  improves the efficiency of these devices. 
 \end{abstract}

\maketitle


In recent years, organic photovoltaics (OPVs) have been the subject of numerous studies due to their potential as an inexpensive and efficient energy source. Power conversion efficiencies of these devices have improved dramatically, now exceeding 10\%.\cite{eff_imp_full_lay,sim_enh_op_circ,tand_pol_sol_cell,sol_cell_eff_tab,fact_lim_div_eff,Nelson}. Although numerous structures have been studied, devices based on semiconducting fullerene derivatives as acceptors dominate the research. In particular, PCBM (formally: 1-(3-methoxycarbonyl)propyl-1-phenyl[6,6]C$_{61}$) is frequently used in the fabrication of OPVs.\cite{high_efficiency,Nelson} Since the discovery that cesium carbonate (Cs$_2$CO$_3$) interlayers could improve the efficiency of inverted OPVs\cite{eff_inv_pol_sol}, interlayers of cesium salts such as cesium carbonate\cite{low_work_fun_surf,dip_ind_anom_s,high_eff_inv_pol}, cesium iodide \cite{Teng}, cesium fluoride\cite{inv_org_sol_cel}, cesium acetate\cite{eff_pol_bulk_het} and cesium stearate\cite{high_eff_org_pho} have become common architectural features of fullerene-based devices. Recent X-ray near-total-reflection fluorescence (XNTRF) studies of devices containing CsI, P3HT (formally: poly(3-hexylthiophene-25-diyl)) and PCBM in a CsI/P3HT:PCBM structure have demonstrated the diffusion of cesium ions into the P3HT:PCBM layer.\cite{self} Secondary ion mass spectrometry (SIMS) profiles of CsF-containing devices also demonstrated the diffusion of cesium into the P3HT:PCBM layer\cite{inv_org_sol_cel}, and the concentration profile of Cs in this layer qualitatively matches the concentration profile of PCBM in a similar layer, as determined by a different, neutron-reflectivity study.\cite{nano_conc_prof}

These studies suggest a strong interaction between PCBM and cesium ions. This may be due to the fact that PCBM is an acceptor and is very electronegative, whereas cesium is an electropositive ion. Indeed, intercalation studies have previously demonstrated the interaction of cesium with carbon compounds.\cite{int_comp_of_graph} Although the interaction of cesium and fullerenes may have a significant effect on the efficiency of cesium containing OPVs, no study has demonstrated the nature of the effect cesium ions have on PCBM. Here, we examine the effect of  cesium carbonate on PCBM  in the form of films at the vapor/water interface.  Fullerene compounds - including PCBM - are highly hydrophobic.  It is therefore possible to study their behavior by spreading them on the surface of water. Once they are spread, it is possible to manipulate the surface density by compression in a Langmuir trough. Whereas pure C$_{60}$ does not form a single continuous layer at the air water interface due to strong aggregation forces, highly functionalized fullerene derivatives can form monomolecular layers\cite{c60_prop_add_mon} and even exhibit short range in-plane order at the vapor/water interface\cite{Fukuto1997}. Pure PCBM, which has a single functional group, is expected to behave more like pure C$_{60}$. In the present study, we examine by X-ray reflectivity and XNTRF techniques the structure of films of PCBM that are spread from chloroform:methanol solutions and compare them with films of a similar solution containing PCBM and Cs$_2$CO$_3$ (known to be highly soluble in both water and alcohols)  Our intention in studying pure PCBM film on water is to examine the interactions and effect Cs$_2$CO$_3$ has on PCBM that may clarify the role the salt plays in improving the performance of inverted OPV devices.

PCBM was purchased from Nano-C and was used as obtained. Chloroform and methanol were purchased from Fisher Scientific. Cs$_2$CO$_3$ (99\% purity) was purchased from Sigma Aldrich and was used as obtained. To prepare each solution, a stock solution of 10 g/L PCBM in chloroform was produced. 250 $\mu$L of PCBM stock solution was then diluted in 10 mL of $\sim$1:1 chloroform:methanol. Finally, 80 mg of Cs$_2$CO$_3$ was added to one of the dilute PCBM solutions. The result was a solution of 0.25 g/L PCBM and 8.0 g/L Cs$_2$CO$_3$ (along with a control solution of 0.25 g/L PCBM) in 1:1 chloroform:methanol. To perform reflectivity measurements, these solutions were spread on ultrapure water (Millipore, Milli-Q, 18.2 M$\Omega$-cm) using a 100 $\mu$L pipette. Here the majority of the Cs$_2$CO$_3$ dissolved into the water, whereas the hydrophobic PCBM remained at the surface.
To compress and characterize PCBM films, a water subphase was maintained at a constant 20$^\circ$C in a thermostatic, Teflon Langmuir trough 

PCBM solutions were spread on the water surface, and the surface pressure ($\Pi$) was measured using a filter-paper Wilhelmy plate. Since the surface concentration of PCBM solution was so dilute before compression, it was expected that Cs$_2$CO$_3$ would rapidly dissolve into the aqueous subphase. According to this hypothesis, without a strong interaction between Cs$_2$CO$_3$ and PCBM,  only PCBM would remain at the surface by the time measurements were taken. Prior to compression, 700 $\mu$L of pure PCBM solution and 500 $\mu$L of PCBM/Cs$_2$CO$_3$ solution were spread. Each solution was allowed to equilibrate for 30 minutes, and was then compressed until a pressure of $\Pi\approx9$ mN/m was reached. 

 The trough for X-ray measurements was sealed in an airtight container and water-saturated helium was flushed through the trough. This protected the sample from radiation damage, decreased diffuse-scattering from air. Reflectivity measurements were performed on the Ames Laboratory liquid surface spectrometer using an UltraX-18 Rigaku X-ray source (rotating copper anode) operating at 50 kV and 250 mA. 
 . A Ge(111) crystal selected the Cu K$\alpha$ wavelength ($\lambda=1.54\AA$), steering the downstream beam towards the liquid surface at a desired angle of incidence $\alpha_i$.

Using a discrete multi-layer model (sometimes known as a multi-box model), electron number density ($\rho_e$) profiles of the liquid surface are used to calculate the expected reflectivities from the films. The calculated reflectivity ($R$) at a given momentum transfer ($Q_z = 4\pi\sin(\alpha_i)/\lambda$) is given by  
\begin{equation}
R(Q_z)=R_0(Q_z)e^{-Q_z^2\sigma^2}
\label{EQ:1}
\end{equation}
where $Q_z$ is along the surface normal ($z$-axis), $\sigma$ is the surface roughness and $R_0(Q_z)$ is the reflectivity of the intrinsic electron density profile in the absence of surface roughness ($\sigma=0$). $R_0(Q_z)$ is calculated using the Paratt's exact recursive method, assuming a 2-box model for the electron density.\cite{el_mod_x_phys,surf_stud_sol_tot} Electron density and roughness parameters are calculated using non-linear least-squares curve-fitting methods. 

This model assumes that the electron-density of each layer is uniform. To a first approximation, severe roughness of the layer substantially reduces $\rho_e$, since voids in the layer artificially decrease the fit $\rho_e$ value. In such cases, the parameters which are generally used to define a film, $\rho_e$, $\sigma$ and $\Delta$ (film thickness), are coupled and it is practically impossible to disentangle them. For X-ray fluorescence measurements, a Vortex-EX multicathode X-ray energy dispersive detector (EDD) is suspended approximately 2 cm above the liquid surface. Fluorescence intensity is recorded as a function of the incident angle. Estimates of the ion surface densities at the surface have been detailed elsewhere.\cite{vaknin} Fluorescence below the critical angle of water ($\alpha_c = 0.152^\circ$, the angle where $Q_c\approx 0.022$ {\AA}$^{-1}$) is almost entirely from ions at the surface (up to a depth of $\sim80$ $\AA$ near the critical angle), whereas the fluorescence above the critical angle includes contributions from both the surface and the subphase\cite{wang2011ionic}. 
 
\begin{figure}[!b]
\includegraphics[width=2.6 in]{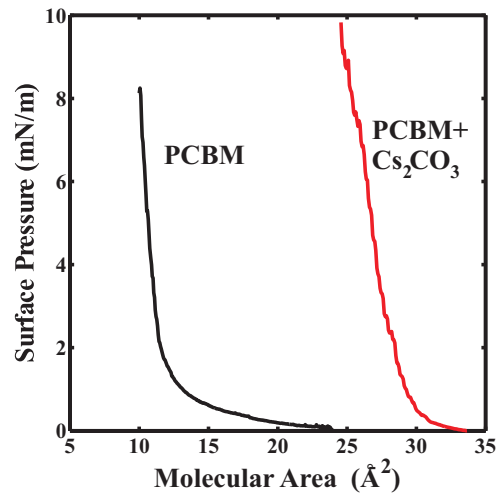}
\caption{(Color online) Area-pressure isotherms for films of pure PCBM and PCBM/Cs$_2$CO$_3$  at the vapor/water interface as indicated. Molecular area is calculated per PCBM molecule for both curves. XR and XNTRF analysis was performed at the highest surface pressure ($\Pi$) of each isotherm. Uncertainty in $\Pi$ is approximately $\pm 1$ mN/m.}
\label{fig:isotherm}
\end{figure}
Figure \ref{fig:isotherm} shows surface pressure versus PCBM-molecular area isotherms of pure PCBM (black) and PCBM/Cs$_2$CO$_3$ (red) solutions. Both isotherms show that the molecular area is much smaller than expected from a cross section of PCBM estimated at $\approx 100$ {\AA}$^2$.  This is strong evidence that the films formed on the surface of water consist of multi-layered PCBM molecules. Based on the cross section of a PCBM molecule ($\sim 100$ {\AA}$^2$) we estimate that the average thickness of the film is 10 or 4 layers of PCBM, depending on whether PCBM was spread from  pure solvent or was mixed with Cs$_2$CO$_3$, respectively.  These isotherms - reproduced on two different Langmuir troughs - suggest that the addition of cesium carbonate has a significant effect on the organization of PCBM films on  water surfaces.   XR and XNTRF measurements were made at the highest observed pressure for each isotherm.

\begin{figure}
\includegraphics[width=2.6 in]{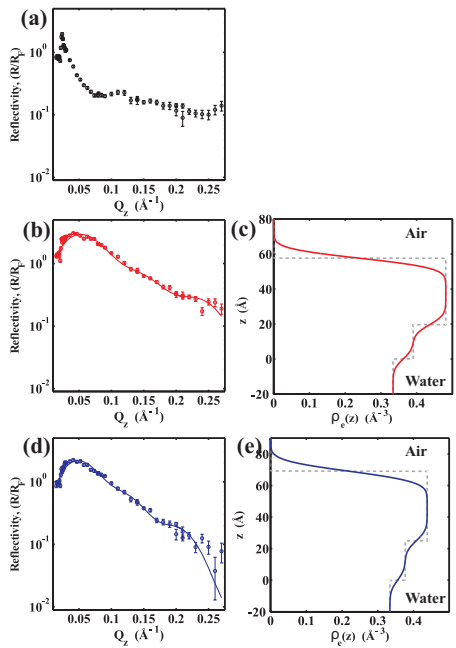}
\caption{ (A), (B) and (D): Respectively, the reflectivity profiles for pure PCBM on water, PCBM/Cs$_2$CO$_3$ on water immediately after spreading and PCBM/Cs$_2$CO$_3$ on water 2 days after spreading. (C) and (E): Electron density profiles corresponding to (B) and (D), respectively. The electron densities were fitted using 2-box models. Due to the high roughness of pure PCBM (see text), these data were not fit. (Color online)} 
\label{fig:ED}
\end{figure}
Normalized X-ray reflectivities ($R/R_F$, where $R_F$ is the Fresnel reflectivity of ideally flat water surface) of PCBM films spread from pure and cesium carbonate solutions are displayed in Figure \ref{fig:ED}. The measurements were conducted at the highest stable surface pressures achievable for both films, as shown in Figure \ref{fig:isotherm}. Qualitative examination of Figure \ref{fig:ED}A of the pure PCBM shows dramatic fall of $R/R_F$ above  $Q_c$, indicating an extremely rough surface. This profile is similar to reflectivities from pure C$_{60}$ molecule at the vapor/water interface.\cite{Vaknin1995} In contrast, above $Q_c$ the $R/R_F$ from PCBM mixed cesium carbonate is more than an order of magnitude stronger (Figure \ref{fig:ED}C), indicating a much smoother and more uniform film.  We anticipate that a reflectivity with an intensity as low as that of pure PCBM would be impossible to fit using the equations developed to model uniform strata. However, using Equation \ref{EQ:1} we can estimate the length scale of the roughness in this film by noting that the reflectivity is lowered by almost a factor of 10 with respect to a smooth film, so that $R/R_F$ is 0.1 at an average $<Q_z>=0.07$ {\AA}$^{-1}$. From Equation \ref{EQ:1} we can estimate the average roughness as $<\sigma> \approx \sqrt{- \ln (R/R_F)/<Q_z>^2}$ so that $<\sigma> \approx 22\AA$. This value is almost twice the diameter of PCBM, implying the density extracted from the reflectivity would be lower than that of densely-packed PCBM. We therefore conclude that analyzing the reflectivity in the standard methods that are suitable for lamellar structures is impractical.  Measurements were recorded for the cesium carbonate solution immediately following spreading and compressing and also approximately 2 days later. The solid lines through the measured reflectivity data in Figure \ref{fig:ED} are best fit  calculations using the model electron densities beside the reflectivity profiles.  Best fit parameters and their uncertainties are listed in Table \ref{tab:ED}.

\begin{table}[!t]
\caption{Best fit parameters (and uncertainties) for reflectivity data from solutions of PCBM. $\rho_e$ is the electron density (\AA$^{-3}$) of a layer, $\Delta$ is the thickness (\AA) of a layer, $\sigma$ (in \AA) is the roughness (assumed to be constant for each interface) and $\chi_{r}^2$ is a measure of the quality of fit, as described in the \emph{Materials and Methods} section. Superscripts 1 and 2 denote (for $\rho_e$ and $\Delta$ values) the box which each parameter represents. Box 1 is the layer closest to air, and box 2 is the layer closest to water.  $^*$The roughness of pure PCBM was estimated in the text. Because this value is so large, applying standard XR analysis is impractical.}
\begin{tabular}{|l|l|l|l|l|l|l|}
\hline
\textbf{Parameters}&\textbf{$\chi_{r}^2$}&{$\rho_e^1$ }&{$\rho_e^2$}&{$\Delta^1$}&{$\Delta^2$}&{$\sigma$} \\[3pt] \hline
{Pure}&&&&&&22{*}\\ \hline
{Cs$_2$CO$_3$}&4.82&0.48$\pm$.02&0.39$\pm$.01&38$\pm$4&20$\pm$4&4.4$\pm$.2\\\hline
{Cs$_2$CO$_3$ (Aged)}&2.67&0.43$\pm$.01&0.37$\pm$.01&44$\pm$5&25$\pm$5&5.4$\pm$.2\\\hline
\end{tabular}
\label{tab:ED}
\end{table}

\begin{figure}[!t]
\includegraphics[width=2.7 in]{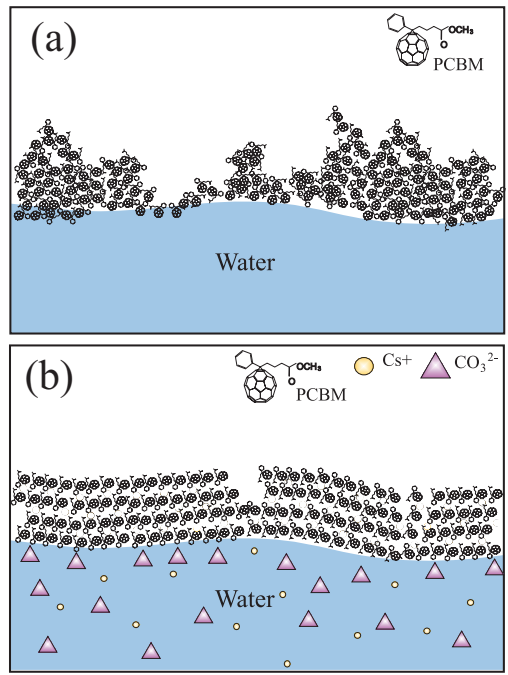}
\caption{A qualitative representation of the difference between A) Pure PCBM on water and B) PCBM/Cs$_2$CO$_3$ solution on water. The films are much smoother when spread from the mixed solution. Although the illustration implies long-range ordering of PCBM/carbonate film, this was not observed experimentally. In fact, synchrotron studies of simliar films have demonstrated only short-range order in both PCBM and PCBM/Cs$_2$CO$_3$.}
\label{fig:results}
\end{figure}

Although the best fit for pure PCBM disagrees with observed reflectivity at very low and very high $Q_z$ (see Figure \ref{fig:ED}) the electron density ($\rho_e$) profile is suggestive of a thick, albeit very rough, surface film. The expected $\rho_e$ of densely packed PCBM is $\approx 0.45$ {\AA}$^{-3}$.\cite{Geens2004,Treat2011,ED_PCBM}, so the reduced $\rho_e$ shown in Figure \ref{fig:ED} A indicates  the formation of voids, as depicted in Figure \ref{fig:results}A. The electron density of the PCBM/Cs$_2$CO$_3$ solution prior to aging ($\rho_e=0.48\AA^{-3}$) exceeds both the observed and the theoretical electron density of closely packed PCBM, suggesting the presence of relatively small amount of cesium in the multilayer.\cite{ED_PCBM} Over time, the electron density of the film decreases slightly  accompanied by  increased  roughness, indicating instability of the film with time. 
\begin{figure}[!b]
\includegraphics[width=2.7 in]{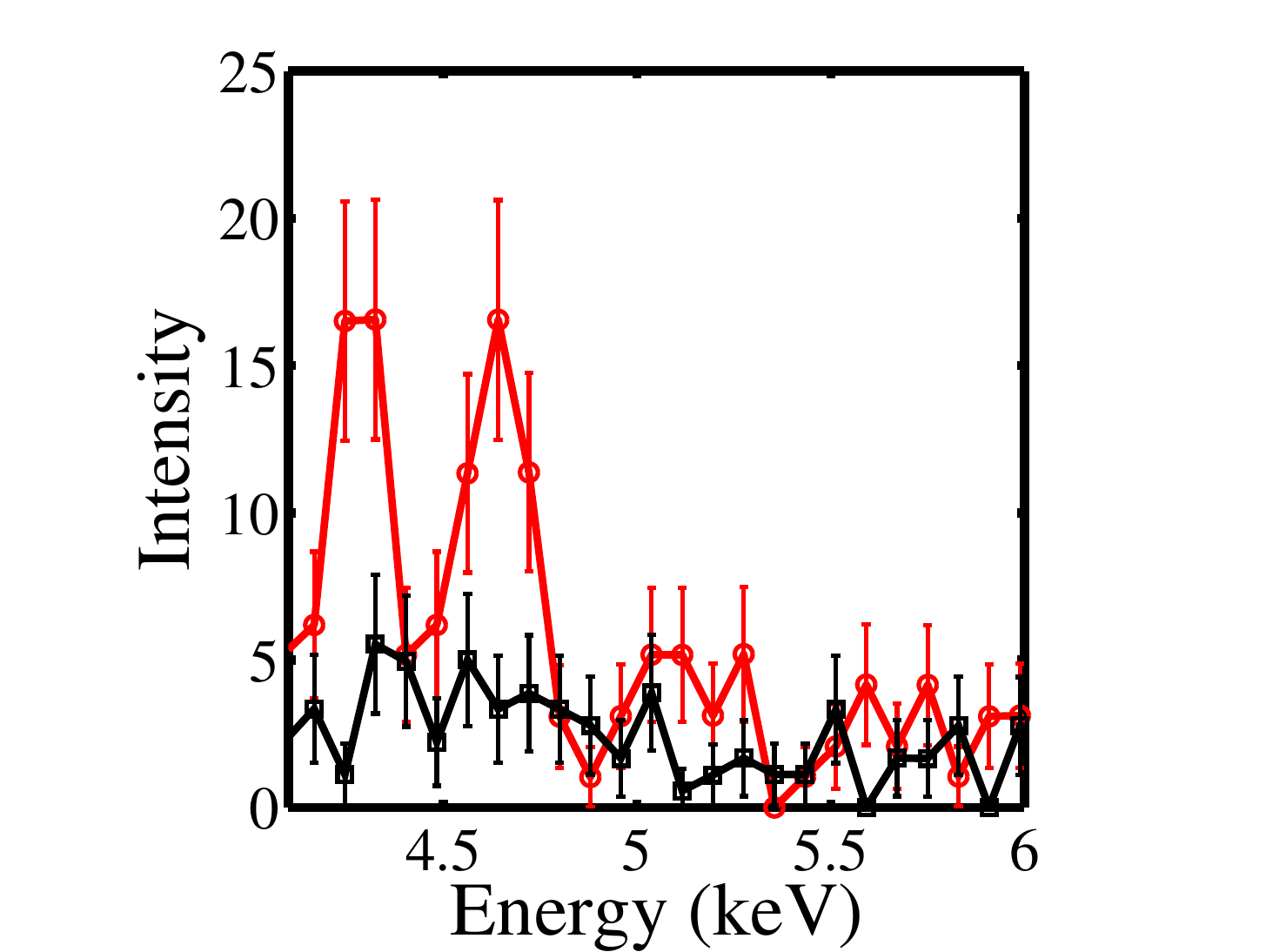}
\caption{(Color online) Fluorescence intensity for 9$\times10^{-5}$M Cs$_2$CO$_3$ subphase (black) and a surface film of PCBM/Cs$_2$CO$_3$ (red) below the critical angle of water. Signal at 4.3 keV corresponds to the L$\alpha_1$ and the L$\alpha_2$ emissions of cesium and signal at 4.6 keV corresponds to the L$\beta_1$ emission. The enhancement of these peaks in the PCBM sample may indicate the presence of cesium in the PCBM layer. Units of intensity are arbitrary.}
\label{fig:Flores}
\end{figure}

Figure \ref{fig:Flores} shows the  fluorescence intensity for 9$\times10^{-5}$M Cs$_2$CO$_3$ subphase (black) and a surface film of PCBM/Cs$_2$CO$_3$ (red) below the critical angle for total reflectivity of water (mentioned above). The signal at 4.3 keV corresponds to the L$\alpha_1$ and the L$\alpha_2$ emissions of cesium and signal at 4.6 keV corresponds to the L$\beta_1$ emission.  Although cesium fluorescence slightly exceeds the detection limit of the EDD, the difference between the integrated surface signal (Figure \ref{fig:Flores}) of the subphase and the solution is still pronounced to establish surface enrichment of Cs compared to its presence in the bulk. Note,  that the fluorescence intensity of the subphase is divided by two, since the expected bulk concentration of Cs$_2$CO$_3$ from the PCBM solution is only 4$\times10^{-5}$M (assuming all the Cs$_2$CO$_3$ of the spread mixture is dissolved in water). This result supports the finding of cesium in the PCBM layer although in a very small amount, namely much lower than one Cs per PCBM, consistent with the X-ray reflectivity analysis.  Similar signal enhancement has been observed in charged Langmuir monolayers that are spread on a variety of ionic solutions (fluorescence signal from ion enriched interfaces has been demonstrated and explained in detail elsewhere\cite{wang2011ionic}). 

In conclusion, X-ray reflectivity results indicate that the introduction of Cs$_2$CO$_3$ into a solution of PCBM in 1:1 chloroform:methanol drastically alters the morphology of PCBM films that are subsequently formed at the vapor/water interface. In particular, the films that are formed with the Cs$_2$CO$_3$ are smoother, with significantly lower roughness than films of pure PCBM.  This  indicates that the cesium carbonate lowers the aggregation forces between PCBM molecules, since these forces would tend to make the film highly nonuniform. This result supports the hypothesis that cesium migration in organic photovoltaics is caused by cesium-PCBM interactions. Based on the relatively low fluorescence intensity of cesium, we estimate that the concentration of cesium within the PCBM film was relatively low - far lower than 1 cesium ion per PCBM molecule - and therefore believe that this system is a reasonable proxy for the behavior of PCBM in OPVs. The addition of cesium salts may therefore have a significant effect on the aggregation of PCBM within OPVs, which may in turn affect  device performance. It is also possible that the interaction between Cs and PCBM can explain the changes in electronic structure reported elsewhere.\cite{Teng}
Similar experiments were attempted on other cesium salt solutions, but most cesium salts could not dissolve in the organic solvents used to dissolve PCBM. Salts containing carbonate ions were also tested, but no suitably soluble carbonate was found. The interaction of cesium carbonate with pure C$_{60}$ was also studied, but finding a solvent for both C$_{60}$ and Cs$_2$CO$_3$ was  difficult. Although fluorescence results indicate the presence of low levels of cesium in the PCBM layer at the vapor/water interface, we cannot rule out the effect of the carbonate on the  aggregation behavior of PCBM in films. Distinguishing the roles of the cesium ions and carbonate ions is difficult because we could not identify other cesium or carbonate containing salts that dissolved in PCBM solutions.

We hypothesize that the addition of Cs$_2$CO$_3$ to OPVs is related to the miscibility of P3HT and PCBM thereby affecting device morphology. Regions of each phase must be large enough so they can percolate to allow holes and electrons to reach the electrodes before recombining and yet small enough domains are necessary to allow a large interfacial area between the two phases, maximizing the dissociation of excitons. Maximizing efficiency means finding an appropriately optimized middle-ground.  Thus, the inclusion of Cs in the vicinity of PCBM may create an optimized network of P3HT/PCBM by lowering the strong aggregation forces between PCBM molecules. 

{\bf Acknowledgements:} We thank Eric Grieser for help with experiments. The work at Ames Laboratory was supported by the Office of Basic Energy Sciences, U.S. Department of Energy under Contract No. DE-AC02-07CH11358, and by the U.S. Department of Energy Science Undergraduate Laboratory Internship (SULI) Program under its contract with Iowa State University, Contract No. DE-AC02-07CH11358. William Lindemann is grateful to the DOE for the assistantship and opportunity to participate in the SULI program. R. Shinar acknowledges partial support of the Iowa Energy Center. 

\bibliographystyle{aipnum4-1}
\bibliography{mybib}

\end{document}